\documentclass[article,nojss,nofooter]{jss}

\usepackage{hyperref}
\usepackage{orcidlink,thumbpdf,lmodern}
\usepackage{bookmark}

\usepackage{amsmath,amssymb}
\usepackage{booktabs}
\usepackage{multirow}
\usepackage{array}

\newcommand{\class}[1]{`\code{#1}'}
\newcommand{\fct}[1]{\code{#1()}}

\author{Nikolaos Korfiatis\thanks{Department of Informatics, Ionian University, Plateia Iatrou Tsirigoti 7, GR-49100 Corfu, Greece. E-mail: \email{nkorf@ionio.gr}}~\orcidlink{0000-0001-6377-4837}\\Ionian University}
\Plainauthor{Nikolaos Korfiatis}

\title{\pkg{grangersearch}: An \proglang{R} Package for Exhaustive Granger Causality Testing with Tidyverse Integration}
\Plaintitle{grangersearch: An R Package for Exhaustive Granger Causality Testing with Tidyverse Integration}
\Shorttitle{\pkg{grangersearch}: Granger Causality in \proglang{R}}

\Abstract{
This paper introduces \pkg{grangersearch}, an \proglang{R} package for performing exhaustive Granger causality searches on multiple time series. The package provides: (1)~exhaustive pairwise search across multiple variables, (2)~automatic lag order optimization with visualization, (3)~tidyverse-compatible syntax with pipe operators and non-standard evaluation, and (4)~integration with the \pkg{broom} ecosystem through \fct{tidy} and \fct{glance} methods. The package wraps the \pkg{vars} infrastructure while providing a simple interface for exploratory causal analysis. We describe the statistical methodology, demonstrate the package through worked examples, and discuss practical considerations for applied researchers.
}

\Keywords{Granger causality, time series analysis, VAR models, exhaustive search, tidyverse, \proglang{R}}
\Plainkeywords{Granger causality, time series analysis, VAR models, exhaustive search, tidyverse, R}

\Address{
  Nikolaos Korfiatis\\
  Department of Informatics\\
  School of Information Science and Informatics\\
  Ionian University\\
  Plateia Iatrou Tsirigoti 7, GR-49100\\
  Corfu, Greece\\
  E-mail: \email{nkorf@ionio.gr}\\
  URL: \url{https://github.com/nkorf/grangersearch}
}

\begin{document}

%% -- Introduction -------------------------------------------------------------

\section[Introduction: Granger causality in R]{Introduction: Granger causality in \proglang{R}} \label{sec:intro}

Understanding causal relationships between time series variables is a fundamental problem in economics, finance, neuroscience, and many other fields. While true causality is philosophically complex and difficult to establish from observational data alone, \citet{granger1969investigating} proposed a practical, testable notion of causality based on predictability: a variable $X$ is said to ``Granger-cause'' another variable $Y$ if past values of $X$ contain information that helps predict $Y$ beyond what is contained in past values of $Y$ alone.

Granger causality testing has found applications across diverse domains. In macroeconomics, \citet{sims1972money} famously applied the technique to study money-income relationships, while \citet{kraft1978relationship} pioneered its use in energy economics. Financial market researchers including \citet{hiemstra1994testing} have extended the methodology to study price-volume dynamics, and neuroscientists have adapted Granger causality for brain connectivity analysis \citep{seth2015granger}. The statistical foundations rest on vector autoregressive (VAR) models \citep{sims1980macroeconomics}, with comprehensive treatments available in \citet{lutkepohl2005new} and discussions of causal interpretation in \citet{peters2017elements}.

Despite its popularity, implementing Granger causality tests in \proglang{R} \citep{R} remains cumbersome for applied researchers. The primary infrastructure is provided by the \pkg{vars} package \citep{vars}, which offers comprehensive VAR modeling capabilities through functions like \fct{VAR} and \fct{causality}. However, users must manually specify variable pairs, construct appropriate model specifications, and interpret raw output. The \pkg{lmtest} package \citep{lmtest} provides a simpler \fct{grangertest} function for bivariate analysis, but it lacks multi-variable search capabilities. The \pkg{MSBVAR} package offers \fct{granger.test} for producing matrices of all pairwise tests, but development has stalled. More recently, the \pkg{bruceR} package has added \fct{granger\_causality} for multivariate testing within a broader toolkit, and the \pkg{NlinTS} package \citep{NlinTS} extends Granger causality to nonlinear settings using neural networks. None of these solutions, however, integrate seamlessly with the tidyverse ecosystem \citep{wickham2019welcome} that has become central to modern \proglang{R} workflows.

This paper presents \pkg{grangersearch}, an \proglang{R} package that wraps the \pkg{vars} infrastructure while providing:
\begin{itemize}
  \item A single function interface that tests causality in both directions
  \item Exhaustive pairwise search across multiple time series variables
  \item Automatic lag order optimization with visualization
  \item Tidyverse-compatible syntax with pipe operators and non-standard evaluation
  \item Structured output objects with all relevant test statistics
  \item Integration with the \pkg{broom} package \citep{robinson2014broom} ecosystem
\end{itemize}

The remainder of this paper is organized as follows. Section~\ref{sec:methodology} reviews the statistical methodology underlying Granger causality tests. Section~\ref{sec:software} compares existing \proglang{R} software for Granger causality analysis. Section~\ref{sec:package} describes the design and functionality of \pkg{grangersearch}. Section~\ref{sec:examples} provides worked examples using the Canadian macroeconomic dataset from \pkg{vars}. Section~\ref{sec:discussion} discusses practical considerations, limitations, and potential extensions. Section~\ref{sec:summary} concludes.

%% -- Methodology --------------------------------------------------------------

\section{Statistical methodology} \label{sec:methodology}

\subsection{Granger causality}

The concept of Granger causality is based on two principles: (1) the cause occurs before the effect, and (2) the cause contains unique information about the effect that is not available elsewhere \citep{granger1969investigating}.

Formally, let $\{X_t\}$ and $\{Y_t\}$ be two stationary time series. We say that $X$ Granger-causes $Y$ if:
\begin{equation} \label{eq:granger}
\sigma^2(Y_t | Y_{t-1}, Y_{t-2}, \ldots) > \sigma^2(Y_t | Y_{t-1}, Y_{t-2}, \ldots, X_{t-1}, X_{t-2}, \ldots)
\end{equation}
where $\sigma^2(Y_t | \cdot)$ denotes the variance of the optimal linear predictor of $Y_t$ given the conditioning information. In other words, $X$ Granger-causes $Y$ if including past values of $X$ improves the prediction of $Y$ beyond what is achieved using past values of $Y$ alone.

\subsection{Vector autoregressive models}

Testing for Granger causality is typically implemented using vector autoregressive (VAR) models \citep{sims1980macroeconomics}. A bivariate VAR model of order $p$, denoted VAR($p$), can be written as:
\begin{eqnarray}
Y_t &=& \alpha_1 + \sum_{i=1}^{p} \beta_{1i} Y_{t-i} + \sum_{i=1}^{p} \gamma_{1i} X_{t-i} + \varepsilon_{1t} \label{eq:var1} \\
X_t &=& \alpha_2 + \sum_{i=1}^{p} \beta_{2i} X_{t-i} + \sum_{i=1}^{p} \gamma_{2i} Y_{t-i} + \varepsilon_{2t} \label{eq:var2}
\end{eqnarray}
where $\alpha_1, \alpha_2$ are intercepts, $\beta_{ji}$ and $\gamma_{ji}$ are autoregressive coefficients, and $\varepsilon_{1t}, \varepsilon_{2t}$ are white noise error terms that may exhibit contemporaneous correlation. The errors are assumed to satisfy $\E[\varepsilon_{jt}] = 0$, $\E[\varepsilon_{jt}\varepsilon_{js}] = 0$ for $t \neq s$, and $\E[\varepsilon_{jt}\varepsilon_{kt}] = \sigma_{jk}$ (possibly non-zero).

\subsection{Hypothesis testing}

Testing whether $X$ Granger-causes $Y$ reduces to testing whether the coefficients $\gamma_{1i}$ in Equation~\ref{eq:var1} are jointly zero:
\begin{equation} \label{eq:null}
H_0: \gamma_{11} = \gamma_{12} = \cdots = \gamma_{1p} = 0
\end{equation}
Under the null hypothesis, the lagged values of $X$ provide no additional predictive information about $Y$ beyond what is contained in the lagged values of $Y$ itself.

This hypothesis can be tested using a standard $F$-test comparing the restricted model (excluding $X$ lags) to the unrestricted model (including $X$ lags). The test statistic is:
\begin{equation} \label{eq:fstat}
F = \frac{(\text{RSS}_R - \text{RSS}_U)/p}{\text{RSS}_U/(T - 2p - 1)}
\end{equation}
where RSS$_R$ and RSS$_U$ denote the residual sum of squares from the restricted and unrestricted models respectively, $T$ is the sample size, and $p$ is the lag order. Under $H_0$, this statistic follows an $F(p, T - 2p - 1)$ distribution asymptotically.

The \pkg{grangersearch} package performs this test in both directions---testing whether $X$ Granger-causes $Y$ and whether $Y$ Granger-causes $X$---providing a complete picture of the predictive relationships between two series.

\subsection{Lag order selection}

The choice of lag order $p$ in the VAR specification is critical and can substantially influence test results. Selecting too few lags may fail to capture the true dynamic relationship, while selecting too many reduces statistical power and may introduce spurious correlations due to estimation of unnecessary parameters.

Information criteria provide data-driven guidance for lag selection. The most commonly used criteria are the Akaike Information Criterion (AIC) and the Bayesian Information Criterion (BIC):
\begin{eqnarray}
\text{AIC}(p) &=& \log|\hat{\Sigma}_p| + \frac{2pK^2}{T} \\
\text{BIC}(p) &=& \log|\hat{\Sigma}_p| + \frac{pK^2 \log(T)}{T}
\end{eqnarray}
where $\hat{\Sigma}_p$ is the estimated residual covariance matrix from the VAR($p$) model, $K$ is the number of variables, and $T$ is the sample size. BIC tends to favor more parsimonious models due to its heavier penalty term, while AIC may select larger lag orders. The \pkg{vars} package provides \fct{VARselect} for computing these criteria.

An alternative approach, implemented in \pkg{grangersearch}, is to examine the sensitivity of Granger causality test results across multiple lag specifications. Results that remain significant across a range of reasonable lag orders provide stronger evidence for robust causal relationships \citep{toda1995statistical}.

\subsection{Stationarity and cointegration}

The standard Granger causality test assumes that the time series under analysis are (weakly) stationary. When series are non-stationary---exhibiting unit roots---test statistics may not follow their assumed distributions, potentially leading to spurious inference.

Researchers should therefore test for stationarity prior to Granger causality analysis. Common approaches include the Augmented Dickey-Fuller (ADF) test and the Kwiatkowski-Phillips-Schmidt-Shin (KPSS) test. Non-stationary series can often be rendered stationary through differencing.

When non-stationary series are cointegrated---sharing a common stochastic trend---a vector error correction model (VECM) framework is more appropriate than a VAR in differences \citep{toda1995statistical}. The \pkg{vars} package provides tools for cointegration analysis through functions like \fct{ca.jo} and \fct{vec2var}. The current version of \pkg{grangersearch} assumes stationarity; handling of cointegrated systems is planned for future releases.

%% -- Software comparison ------------------------------------------------------

\section[Existing software for Granger causality in R]{Existing software for Granger causality in \proglang{R}} \label{sec:software}

Table~\ref{tab:comparison} provides an overview of \proglang{R} packages offering Granger causality testing functionality. Each package represents a different design philosophy and target use case.

\begin{table}[t!]
\centering
\begin{tabular}{lcccccc}
\toprule
Package & Bivariate & Multivar. & Search & Tidyverse & Broom & Active \\
\midrule
\pkg{vars} & \checkmark & \checkmark & --- & --- & --- & \checkmark \\
\pkg{lmtest} & \checkmark & --- & --- & --- & --- & \checkmark \\
\pkg{MSBVAR} & \checkmark & \checkmark & \checkmark & --- & --- & --- \\
\pkg{bruceR} & \checkmark & \checkmark & --- & --- & --- & \checkmark \\
\pkg{NlinTS} & \checkmark & --- & --- & --- & --- & \checkmark \\
\pkg{grangersearch} & \checkmark & --- & \checkmark & \checkmark & \checkmark & \checkmark \\
\bottomrule
\end{tabular}
\caption{\label{tab:comparison} Comparison of \proglang{R} packages for Granger causality testing. ``Multivar.'' indicates support for multivariate (conditional) Granger causality. ``Search'' indicates automatic pairwise search across variables. ``Tidyverse'' indicates pipe compatibility and NSE support. ``Broom'' indicates \fct{tidy}/\fct{glance} methods. ``Active'' indicates ongoing maintenance as of 2024.}
\end{table}

The \pkg{vars} package \citep{vars} provides the most comprehensive VAR modeling infrastructure in \proglang{R}. Its \fct{causality} function implements both Granger and instantaneous causality tests within a fully multivariate framework, conditioning on all other variables in the system. This approach is statistically rigorous but requires users to pre-specify the VAR model and manually interpret the output. The package excels for confirmatory analysis where researchers have specific hypotheses about causal structure.

The \pkg{lmtest} package \citep{lmtest} offers a simpler interface through \fct{grangertest}, which conducts bivariate Granger causality tests using standard linear model machinery. While easy to use, it tests only one direction at a time and provides no multi-variable search capability.

The \pkg{MSBVAR} package historically provided \fct{granger.test} for producing matrices of all pairwise Granger causality tests. This functionality anticipated the exhaustive search approach of \pkg{grangersearch}. However, the package has not been updated since 2015 and is not compatible with recent \proglang{R} versions.

The \pkg{bruceR} package includes \fct{granger\_causality} as part of a broader statistical toolkit. It supports multivariate testing based on VAR models but does not provide automatic search across variable pairs or tidyverse integration.

The \pkg{NlinTS} package extends Granger causality to nonlinear settings using neural network models (VARNN). This addresses an important limitation of standard linear Granger causality but focuses on methodology rather than workflow integration.

The \pkg{grangersearch} package takes a different approach: it combines exhaustive search capabilities with a tidyverse-compatible interface. While existing packages require users to specify variable pairs manually, \pkg{grangersearch} automates the discovery process across multiple variables, making it suitable for exploratory analysis. The integration with tidyverse conventions through pipe operators and non-standard evaluation ensures that the package fits naturally into modern \proglang{R} workflows, while the structured output objects and broom compatibility support reproducible research practices.

%% -- Package description ------------------------------------------------------

\section{Package description} \label{sec:package}

\subsection{Installation}

The \pkg{grangersearch} package can be installed from GitHub using the \pkg{devtools} package \citep{devtools}:
\begin{CodeChunk}
\begin{CodeInput}
R> devtools::install_github("nkorf/grangersearch")
\end{CodeInput}
\end{CodeChunk}
The package depends on \pkg{vars} \citep{vars} for VAR model estimation, \pkg{rlang} \citep{rlang} for tidy evaluation, \pkg{tibble} for tibble output, and \pkg{generics} for S3 method registration.

\subsection{Core function: granger\_causality\_test()}

The primary function for bivariate testing is \fct{granger\_causality\_test}. Its signature is:
\begin{Code}
granger_causality_test(.data = NULL, x, y, lag = 1, alpha = 0.05,
  test = "F")
\end{Code}
The arguments are:
\begin{description}
  \item[\code{.data}] Optional data frame or tibble containing the time series variables.
  \item[\code{x}, \code{y}] Numeric vectors containing the time series, or (if \code{.data} is provided) unquoted column names.
  \item[\code{lag}] Integer specifying the lag order for the VAR model (default: 1).
  \item[\code{alpha}] Significance level for hypothesis testing (default: 0.05).
  \item[\code{test}] Test type; currently only \code{"F"} is supported.
\end{description}

The function returns an S3 object of class \class{granger\_result} containing test results for both directions of causality. Table~\ref{tab:output} describes the components.

\begin{table}[t!]
\centering
\begin{tabular}{lp{9cm}}
\toprule
Component & Description \\
\midrule
\code{x\_causes\_y} & Logical indicating whether $X$ Granger-causes $Y$ at level $\alpha$ \\
\code{y\_causes\_x} & Logical indicating whether $Y$ Granger-causes $X$ at level $\alpha$ \\
\code{p\_value\_xy} & $p$-value for the test of $X \rightarrow Y$ \\
\code{p\_value\_yx} & $p$-value for the test of $Y \rightarrow X$ \\
\code{test\_statistic\_xy} & $F$-statistic for the $X \rightarrow Y$ test \\
\code{test\_statistic\_yx} & $F$-statistic for the $Y \rightarrow X$ test \\
\code{lag} & Lag order used in VAR specification \\
\code{alpha} & Significance level used for testing \\
\code{n} & Number of observations \\
\code{x\_name}, \code{y\_name} & Names of the input variables \\
\bottomrule
\end{tabular}
\caption{\label{tab:output} Components of the \class{granger\_result} object returned by \fct{granger\_causality\_test}.}
\end{table}

\subsection{Tidyverse integration}

A key design goal of \pkg{grangersearch} is seamless integration with the tidyverse ecosystem \citep{wickham2019welcome}. The package supports both the \pkg{magrittr} pipe (\code{\%>\%}) and the native \proglang{R} pipe (\code{|>} introduced in version 4.1.0):
\begin{CodeChunk}
\begin{CodeInput}
R> library("grangersearch")
R> data("Canada", package = "vars")
R>
R> # Using native pipe
R> Canada |> granger_causality_test(e, U, lag = 2)
R>
R> # Using magrittr pipe
R> Canada %>% granger_causality_test(e, U, lag = 2)
\end{CodeInput}
\end{CodeChunk}
Non-standard evaluation (NSE) allows column names to be passed unquoted when using a data frame, making code more readable and consistent with other tidyverse functions:
\begin{CodeChunk}
\begin{CodeInput}
R> # Unquoted column names (with data frame)
R> granger_causality_test(Canada, e, U)
R>
R> # Equivalent explicit specification
R> granger_causality_test(x = Canada$e, y = Canada$U)
\end{CodeInput}
\end{CodeChunk}

The package provides \fct{tidy} and \fct{glance} methods following \pkg{broom} conventions \citep{robinson2014broom}. The \fct{tidy} method returns a tibble with one row per direction of causality:
\begin{CodeChunk}
\begin{CodeInput}
R> result <- Canada |> granger_causality_test(e, U, lag = 2)
R> tidy(result)
\end{CodeInput}
\begin{CodeOutput}
# A tibble: 2 x 6
  direction cause effect statistic   p.value significant
  <chr>     <chr> <chr>      <dbl>     <dbl> <lgl>
1 e -> U    e     U          16.7  0.0000003 TRUE
2 U -> e    U     e           1.23 0.298     FALSE
\end{CodeOutput}
\end{CodeChunk}
The \fct{glance} method returns a single-row tibble summarizing the overall test:
\begin{CodeChunk}
\begin{CodeInput}
R> glance(result)
\end{CodeInput}
\begin{CodeOutput}
# A tibble: 1 x 5
    lag alpha     n x_name y_name
  <dbl> <dbl> <int> <chr>  <chr>
1     2  0.05    84 e      U
\end{CodeOutput}
\end{CodeChunk}

\subsection{Exhaustive search: granger\_search()}

The \fct{granger\_search} function implements exhaustive pairwise Granger causality testing across multiple variables:
\begin{Code}
granger_search(.data, ..., lag = 1, alpha = 0.05, test = "F",
  include_insignificant = FALSE)
\end{Code}
The arguments are:
\begin{description}
  \item[\code{.data}] Data frame or tibble containing the time series variables.
  \item[\code{...}] Optional column selection using tidyselect syntax. If omitted, all numeric columns are tested.
  \item[\code{lag}] Integer lag order, or a vector of lags to search over.
  \item[\code{alpha}] Significance level for filtering results.
  \item[\code{include\_insignificant}] Logical; if \code{FALSE} (default), only significant relationships are returned.
\end{description}

For a dataset with $K$ numeric columns, the function tests all $K(K-1)$ directed pairs and returns results sorted by $p$-value. When a vector of lags is provided (e.g., \code{lag = 1:4}), the function tests each lag order and reports the result with the smallest $p$-value for each pair.

The function returns an S3 object of class \class{granger\_search} with a \fct{plot} method that produces a causality matrix visualization (see Section~\ref{sec:examples} for examples).

\subsection{Lag selection: granger\_lag\_select()}

The \fct{granger\_lag\_select} function systematically evaluates Granger causality tests across multiple lag orders:
\begin{Code}
granger_lag_select(.data = NULL, x, y, lag = 1:4, alpha = 0.05,
  test = "F")
\end{Code}
The function returns an S3 object of class \class{granger\_lag\_select} containing detailed results for each lag order tested. A \fct{plot} method visualizes how $p$-values vary across lag specifications, helping researchers assess the robustness of results and identify the optimal lag order.

\subsection{S3 methods}

The package provides intuitive \fct{print} and \fct{summary} methods for all result objects. For \class{granger\_result}:
\begin{CodeChunk}
\begin{CodeInput}
R> result <- Canada |> granger_causality_test(e, U, lag = 2)
R> print(result)
\end{CodeInput}
\begin{CodeOutput}
Granger Causality Test
======================

Observations: 84, Lag order: 2, Significance level: 0.050

e -> U: e Granger-causes U (p = 0.0000)
U -> e: U does not Granger-cause e (p = 0.2983)
\end{CodeOutput}
\end{CodeChunk}

%% -- Examples -----------------------------------------------------------------

\section{Illustrations} \label{sec:examples}

We illustrate \pkg{grangersearch} using the \code{Canada} dataset from the \pkg{vars} package, which contains quarterly Canadian macroeconomic data from 1980Q1 to 2000Q4. The four variables are: \code{e} (employment), \code{prod} (labor productivity), \code{rw} (real wage), and \code{U} (unemployment rate). This dataset has been widely used in econometrics teaching and research.

\subsection{Basic bivariate testing}

We begin by loading the data and examining the relationship between employment (\code{e}) and unemployment (\code{U}):
\begin{CodeChunk}
\begin{CodeInput}
R> library("grangersearch")
R> data("Canada", package = "vars")
R>
R> # Test with lag order 2
R> result <- Canada |>
+    granger_causality_test(e, U, lag = 2)
R> result
\end{CodeInput}
\begin{CodeOutput}
Granger Causality Test
======================

Observations: 84, Lag order: 2, Significance level: 0.050

e -> U: e Granger-causes U (p = 0.0000)
U -> e: U does not Granger-cause e (p = 0.2983)
\end{CodeOutput}
\end{CodeChunk}
The results indicate a unidirectional Granger-causal relationship: employment significantly predicts unemployment ($p < 0.0001$), but unemployment does not significantly predict employment ($p = 0.30$). This finding is consistent with labor market theory, where changes in employment typically precede changes in unemployment rates.

\subsection{Obtaining tidy output}

The \fct{tidy} method facilitates integration with tidyverse workflows:
\begin{CodeChunk}
\begin{CodeInput}
R> library("dplyr")
R>
R> tidy(result) |>
+    filter(significant) |>
+    select(direction, statistic, p.value)
\end{CodeInput}
\begin{CodeOutput}
# A tibble: 1 x 3
  direction statistic   p.value
  <chr>         <dbl>     <dbl>
1 e -> U         16.7 0.0000003
\end{CodeOutput}
\end{CodeChunk}

\subsection{Exhaustive pairwise search}

To discover all significant Granger-causal relationships in the dataset, we use \fct{granger\_search}:
\begin{CodeChunk}
\begin{CodeInput}
R> search_results <- Canada |>
+    granger_search(lag = 2, alpha = 0.05)
R> search_results
\end{CodeInput}
\begin{CodeOutput}
Granger Causality Search Results
================================

4 variables tested: e, prod, rw, U
12 directed pairs examined at lag order 2
4 significant relationships found (alpha = 0.05)

Results (sorted by p-value):
  cause  effect   p.value  lag  significant
  e      U       0.0000003   2  TRUE
  prod   rw      0.0003      2  TRUE
  e      prod    0.0127      2  TRUE
  rw     U       0.0387      2  TRUE
\end{CodeOutput}
\end{CodeChunk}
The search identifies four significant predictive relationships at the 5\% level. Employment Granger-causes both unemployment and productivity, productivity Granger-causes real wages, and real wages Granger-cause unemployment. These findings suggest a plausible causal chain in the Canadian labor market.

To include all tested pairs regardless of significance:
\begin{CodeChunk}
\begin{CodeInput}
R> Canada |>
+    granger_search(lag = 2, include_insignificant = TRUE) |>
+    tidy() |>
+    head(8)
\end{CodeInput}
\begin{CodeOutput}
# A tibble: 8 x 5
  cause effect   p.value   lag significant
  <chr> <chr>      <dbl> <int> <lgl>
1 e     U      0.0000003     2 TRUE
2 prod  rw     0.000300      2 TRUE
3 e     prod   0.0127        2 TRUE
4 rw    U      0.0387        2 TRUE
5 prod  U      0.0784        2 FALSE
6 U     rw     0.196         2 FALSE
7 rw    prod   0.246         2 FALSE
8 U     e      0.298         2 FALSE
\end{CodeOutput}
\end{CodeChunk}

\subsection{Visualizing the causality matrix}

\begin{figure}[t!]
\centering
\includegraphics[width=0.95\textwidth]{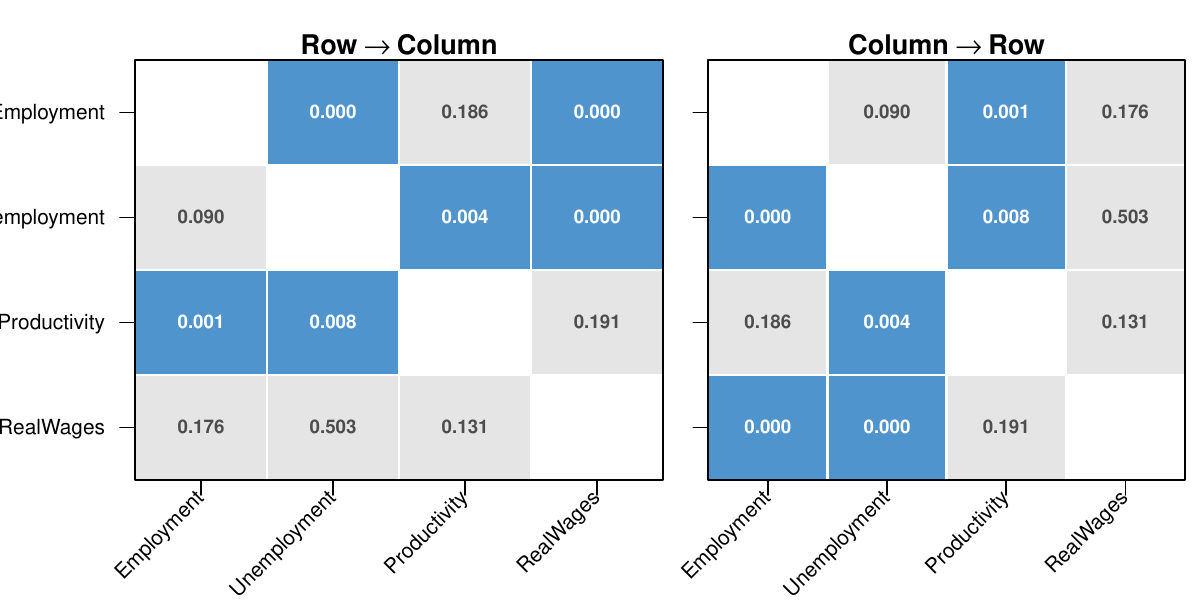}
\caption{\label{fig:causalmatrix} Causality matrix for the Canadian macroeconomic data. Blue cells indicate statistically significant Granger-causal relationships at the 5\% level; gray cells indicate non-significant relationships. The left panel tests whether row variables Granger-cause column variables; the right panel shows the reverse direction.}
\end{figure}

The \fct{plot} method for \class{granger\_search} objects produces a causality matrix visualization:
\begin{CodeChunk}
\begin{CodeInput}
R> plot(search_results)
\end{CodeInput}
\end{CodeChunk}
Figure~\ref{fig:causalmatrix} shows the resulting plot. The left panel displays $p$-values for tests where the row variable Granger-causes the column variable; the right panel shows the reverse direction. Cells are colored blue for significant relationships ($p < 0.05$) and gray otherwise.

\subsection{Lag order selection}

\begin{figure}[t!]
\centering
\includegraphics[width=0.9\textwidth]{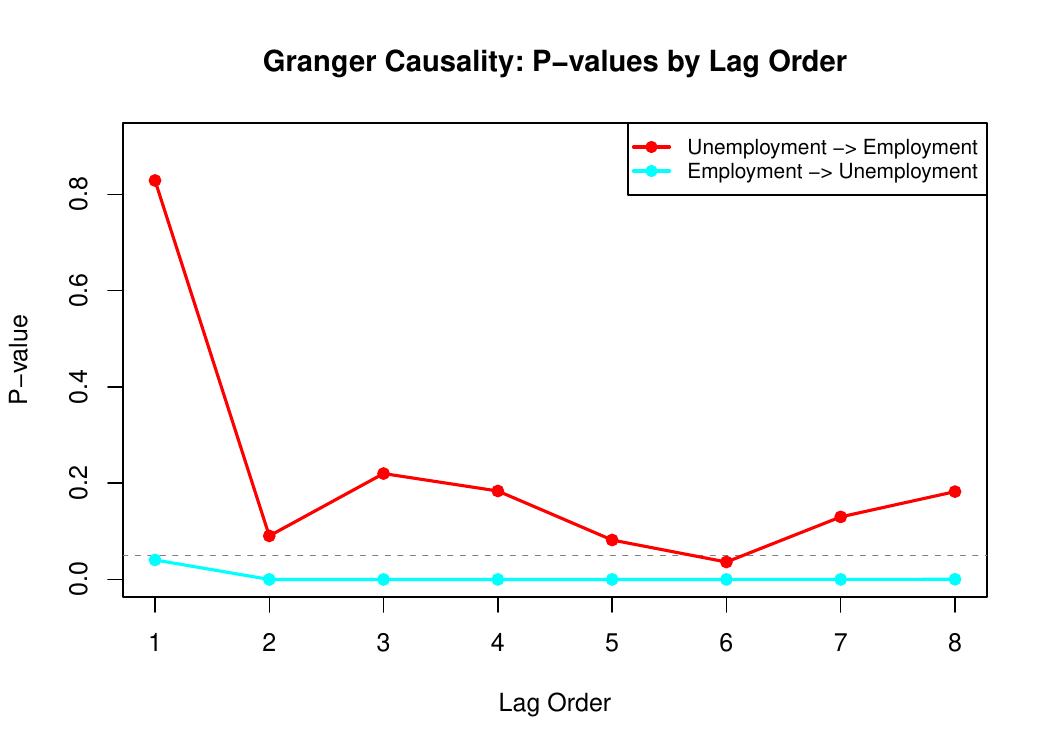}
\caption{\label{fig:lagselect} Lag selection analysis for employment and unemployment. The plot shows $p$-values for both directions of Granger causality across lag orders 1--8. The horizontal dashed line indicates the significance threshold ($\alpha = 0.05$). Employment consistently Granger-causes unemployment regardless of lag specification.}
\end{figure}

To assess the robustness of findings to lag specification, we use \fct{granger\_lag\_select}:
\begin{CodeChunk}
\begin{CodeInput}
R> lag_analysis <- Canada |>
+    granger_lag_select(e, U, lag = 1:8)
R> lag_analysis
\end{CodeInput}
\begin{CodeOutput}
Granger Lag Selection Analysis
==============================

Variables: e -> U (and reverse)
Lag orders tested: 1, 2, 3, 4, 5, 6, 7, 8
Significance level: 0.05

Summary:
  e -> U: Significant at all 8 lag orders
  U -> e: Never significant

Best lag (by minimum p-value):
  e -> U: lag = 2 (p = 0.0000003)
  U -> e: lag = 1 (p = 0.1652)
\end{CodeOutput}
\end{CodeChunk}
The \fct{plot} method visualizes how $p$-values vary across lag orders:
\begin{CodeChunk}
\begin{CodeInput}
R> plot(lag_analysis)
\end{CodeInput}
\end{CodeChunk}
Figure~\ref{fig:lagselect} shows the resulting plot. Employment Granger-causes unemployment at every lag order tested (solid line consistently below the significance threshold), while the reverse relationship is never significant. This consistency strongly supports the robustness of the finding.

\subsection{Searching over multiple lags}

When the optimal lag order is unknown, \fct{granger\_search} can evaluate multiple lags simultaneously:
\begin{CodeChunk}
\begin{CodeInput}
R> Canada |>
+    granger_search(lag = 1:4, alpha = 0.05) |>
+    tidy()
\end{CodeInput}
\begin{CodeOutput}
# A tibble: 4 x 5
  cause effect   p.value   lag significant
  <chr> <chr>      <dbl> <int> <lgl>
1 e     U      0.0000003     2 TRUE
2 prod  rw     0.000300      2 TRUE
3 e     prod   0.0127        2 TRUE
4 rw    U      0.0387        2 TRUE
\end{CodeOutput}
\end{CodeChunk}
When multiple lags are specified, the function reports the lag order that yields the smallest $p$-value for each significant pair.

%% -- Discussion ---------------------------------------------------------------

\section{Discussion} \label{sec:discussion}

\subsection{Practical recommendations}

Applying Granger causality tests correctly requires attention to several issues.

First, researchers should verify that time series are stationary before conducting tests. Unit root tests such as the Augmented Dickey-Fuller (ADF) test or the KPSS test should be applied to each series, and non-stationary series should be differenced appropriately before analysis.

Second, the selection of lag order can substantially influence results. Information criteria such as AIC and BIC provide guidance for lag selection, and the \fct{VARselect} function in \pkg{vars} offers a convenient implementation. The \fct{granger\_lag\_select} function in this package provides an alternative approach by allowing researchers to visualize how test results vary across different lag specifications. Results that remain significant across multiple reasonable lag orders provide stronger evidence.

Third, when conducting exhaustive searches across many variable pairs, multiple testing corrections may be warranted. For $K$ variables, $K(K-1)$ tests are performed. Conservative approaches include Bonferroni correction or controlling the false discovery rate using Benjamini-Hochberg procedures.

Finally, it is important to maintain appropriate caution when interpreting results. Granger causality measures predictive relationships rather than causal mechanisms in the philosophical sense \citep{peters2017elements}. A finding that variable $X$ Granger-causes variable $Y$ indicates that past values of $X$ improve predictions of $Y$, but does not preclude the possibility that both variables are driven by an unobserved common cause.

\subsection{Limitations}

The current implementation has several limitations. The exhaustive search conducts bivariate tests that do not condition on other variables in the dataset. True multivariate Granger causality testing, which conditions on all other variables simultaneously, would provide more robust inference but requires substantially more complex implementation. The \fct{causality} function in \pkg{vars} provides this capability.

The package implements only the $F$-test for hypothesis testing and assumes linear relationships. For nonlinear applications, the \pkg{NlinTS} package offers neural network-based alternatives. The package also assumes that input series are stationary and does not provide built-in handling for cointegrated systems.

\subsection{Future work}

Future versions may include: conditional multivariate testing, automatic stationarity diagnostics with differencing recommendations, nonlinear extensions, network visualization for high-dimensional results, and built-in multiple testing corrections.

%% -- Summary ------------------------------------------------------------------

\section{Summary} \label{sec:summary}

The \pkg{grangersearch} package provides a simple interface for Granger causality testing and discovery in \proglang{R}. By combining statistical rigor with modern \proglang{R} programming practices, the package makes this econometric technique accessible to a broader audience.

Key contributions include: (1) exhaustive pairwise search functionality for automated discovery of causal relationships across multiple variables, (2) lag order optimization tools with visualization for robust model specification, (3) tidyverse integration ensuring the package fits naturally into modern data analysis workflows, and (4) structured output objects compatible with the \pkg{broom} ecosystem.

The package is intended for researchers investigating predictive causal relationships in time series data across economics, finance, and other disciplines where understanding temporal dependencies between variables is of interest.

%% -- Computational details ----------------------------------------------------

\section*{Computational details}

The results in this paper were obtained using \proglang{R}~4.5.1 with packages \pkg{grangersearch}~0.1.0, \pkg{vars}~1.6-1, \pkg{rlang}~1.1.4, \pkg{tibble}~3.2.1, and \pkg{dplyr}~1.1.4. \proglang{R} itself and all packages used are available from the Comprehensive \proglang{R} Archive Network (CRAN) at \url{https://CRAN.R-project.org/}. The \pkg{grangersearch} package is available from GitHub at \url{https://github.com/nkorf/grangersearch}.

\section*{Acknowledgments}

The author thanks the developers of the \pkg{vars} package for providing the underlying VAR modeling infrastructure. The tidyverse ecosystem and \pkg{rlang} package contributed substantially to the design philosophy and implementation of \pkg{grangersearch}.

%% -- Bibliography -------------------------------------------------------------

\bibliography{references}

%% -- Appendix -----------------------------------------------------------------

\newpage

\begin{appendix}

\section{Computational complexity} \label{app:complexity}

The computational cost of exhaustive Granger causality search scales with the number of variables $K$, the number of observations $T$, and the lag order $p$. For each of the $K(K-1)$ directed pairs, a VAR($p$) model must be estimated with $O(T \cdot p^2)$ operations for ordinary least squares. The total complexity is therefore $O(K^2 \cdot T \cdot p^2)$.

For typical applications with $K \leq 20$ variables, $T \leq 1000$ observations, and $p \leq 10$ lags, computation completes in seconds on modern hardware. For larger problems, parallel computation could be implemented in future versions.

\section{Comparison with vars::causality()} \label{app:vars}

The \fct{causality} function in the \pkg{vars} package implements Granger causality testing within a fitted VAR model. Key differences from \pkg{grangersearch} include:

\begin{itemize}
  \item \pkg{vars} requires pre-fitting a VAR model with all variables; \pkg{grangersearch} fits bivariate models for each pair.
  \item \pkg{vars} implements conditional Granger causality (controlling for other variables); \pkg{grangersearch} implements unconditional bivariate tests.
  \item \pkg{vars} tests one pair at a time; \pkg{grangersearch} provides exhaustive search.
  \item \pkg{grangersearch} offers tidyverse integration and \pkg{broom} compatibility.
\end{itemize}

For confirmatory analysis with a pre-specified VAR model, \pkg{vars} is more appropriate. For exploratory analysis discovering potential relationships, \pkg{grangersearch} offers a more streamlined workflow.

\end{appendix}

\end{document}